# Life and Quantum Biology, an Interdisciplinary Approach

Alfred Driessen\*

Summary: *1. Introduction. 2. Life in philosophy. 3. Need of a system approach in science. 4. Discussion and concluding remarks.*

## 1. Introduction

In the last years a number of experimental and theoretical studies have been published which relate quantum mechanics (QM) to biological systems (Vedral 2011), (Romero-Isart et al. 2010), (Collini et al. 2010), (Gerlich et al. 2011), (Panichayangangkoon et al. 2010), (Abbot et al. 2008), (Lloyd 2011), (Ball 2011), (Lambert et al. 2013). The importance of this is not the occurrence of quantum effects in biology – that's nothing new – but the fact that these play a non-trivial role (Davies 2004) as they occur at an unexpected size- and temperature scale. When dealing with QM it seems that one enters a new world, the quantum world, where new concepts are often in conflict with prejudices based on a mechanistic world view of classical physics. For some this new world seems to be strange and exotic, full of, what is called, quantum weirdness. But perhaps part of the strange impressions could be avoided if we would draw more attention to the intellectual tools we are using for our analysis. In the discussion about the fundamental issues of QM inevitably one uses concepts like causality, (un-) determinism, movement, change, reality, space and time: all of these also being studied in philosophy. The foundation of QM belongs not only to the field of science in the sense of natural science. Also philosophy, part of the *Geisteswissenschaften* (humanities) enters this field as it delivers and develops the concepts and tools needed for a fundamental analysis.

There is a big issue with philosophy as, from the outside, one observes a group of people who in nearly 3000 years of history came up with quite different, and sometimes contradictory approaches to explain the ultimate causes of the universe and the beings therein. In spite of this discouraging diversity one can distinguish a classical line of thinking originating from the Greek

\* Emeritus Professor University of Twente, private address: Jan Luijkenstraat 52, 1071 CS Amsterdam, The Netherlands. E-mail: driessen.alfred@gmail.com

The author would like to thank Gerard Nienhuis and Daan van Schalkwijk for helpful discussions.

philosopher Aristotle, further elaborated by Thomas Aquinas and others and continued nowadays as an active field of study at numerous academic institutions. In the following, often reference is made to this specific Aristotelian-Thomistic (A-T) approach, thereby inspired by P. Hoenen, who in his philosophical analysis of modern science (Hoenen 1947) applied in an original way classical concepts. This choice for a philosophy inspired by Aristotle is not arbitrary, because Aristotelian logic, for example, is still at the basis of mathematical logic and mainstream scientific argumentation. Also his analysis of causality and his concept of First Cause is still extensively used, see, for example, Stephen Hawkings's *A brief history of time* (Hawking 1988), (Driessen 1995). Heisenberg, one of the founding fathers of QM, extensively used Aristotelian metaphysics in his explanation of the fundamentals of this new theory (Heisenberg 2007). Soler Gil (2003) demonstrates that the Aristotelian ontology can be successfully applied to objects of QM.

Turning now our attention to living matter, again a new world seems to open, so familiar to us human beings, but also so strange or weird as the quantum world. It is often called the mystery of life. Up to now no biological material or even less a complete cell could be synthesized starting from pure chemical substances. There is a mystery about life in all its forms from the most primitive forms, a virus, a bacteria up to higher levels in flora and fauna. When speaking above about the quantum world it became obvious that scientific terms are also studied or even had their origin in philosophy. The same holds for the answer to the question, "What is life?". Philosophers from the very beginning of ancient Greek history, have developed profound ideas about life, see, e.g., Weber (2011).

For many scientists the ability to understand life and to reproduce biological processes is a matter of time and years of intensive research. In order to achieve this understanding one should unravel the chemical processes and study the underlying physical phenomena in detail. According to this view, biology will eventually become applied chemistry and chemistry applied physics. Others prefer a holistic view instead and consider living matter as something that cannot be reduced to an agglomerate of lifeless building blocks. In engineering a similar situation can be observed. Some concentrate on the components and others prefer a system approach. A unique classification, however, in terms of system or components is not possible, as a multi-level hierarchical structure can be found. This is a general property of complex systems where no natural detail level can be assigned for an adequate description, see, e.g. Mitchell (2009). To give an engineering example, the components of a simple telecommunication system are on their own subsystems, namely a transmitter, receiver and a transmission line. Again, a large number of these telecommunication systems can be integrated to form a communication network or eventually the World Wide Web. Coming back to the highly complex biological systems one may state that reductionism is not the one and only option, see (Brigandt and Love 2012).

In the present study two concepts will be central: information and unity. To begin with the last, unity is the central concept in living matter. This can be seen by considering the opposite, division or decomposition, which is the most clear manifestation of death. In the 19$^{th}$ century mechanism, unity is reduced to a geometrical arrangement of parts, like in a machine. QM has proven that that is not the correct approach, as entanglement and superposition play an essential role. And these new phenomena, which have no classical equivalent, are both closely related to the concept of unity. Information is the other characteristic of living matter as is illustrated, for example, in the title of a paper on the origin of life by the Nobel Laureate Manfred Eigen: *Wie entsteht Information?* (How does information originate) (Eigen 1976). Information is related to the complexity of a system. And living organisms are top-ranked in the ladder of complexity, as they are far away from the boring periodicity of minerals and the trivial chaos in an ideal gas (Arecchi 1993). Information connects also to the philosophy of Aristotle, as the concept of information is related to *the reception or acquisition of a form in the Aristotelian sense* (Del Re 1993).

It is a challenging endeavor to deal with three disciplines: biology, physics and philosophy in a single paper. We will follow the same method Aristotle applied in his work. In ancient Greece it was possible to acquire an adequate knowledge of the science of his time in a single lifetime. Aristotle acquired this knowledge and applied it to lay the foundations of his philosophy and provide illustrative examples. Science since then has made an enormous progress, but the ultimate objective to integrate science and philosophy in a consistent picture still remains. The everyday experience that was sufficient for the Greek philosophers is now increased by scientific evidence that is only understandable by the specialists. Close interaction therefore is needed between philosophers and scientists that eventually would lead to better understanding of philosophy as well as science.

This work starts with an overview of how life is treated in the classical A-T approach and gives also comments on a few alternatives. In the subsequent section the focus is on the relation between the whole and the constituent parts. It is the most challenging part as philosophical concepts and statements are evaluated in the light of scientific experience. The latter mostly include examples from QM and also quantum biology. In the discussion some speculative ideas are presented about life being a kind of macroscopic entangled state (see Vedral 2011) enabling unexpected efficiency in attaining advantageous behavior. Finally a recommendation is given for an open dialogue between philosophers and scientists with an added value for both.

2. Life in philosophy

For common people an animal, for example a dog, is clearly distinguished from non-living objects. Classical philosophy as expressed by Aristotle follows this common-sense distinction and gives the following definition: *by life we mean self-nutrition and growth with its correlative decay* (Aristotle 350 BC). This definition is valid for the three stages of life: plants, animals and human beings. Aquinas adds in his commentary on Aristotle an additional aspect:

«Something is living not only because it has growth and decay, but also because it is able to feel and grasp intellectually and is able to carry out other works of life…It is therefore a property of life to move oneself, whereby movement has to be taken in the widest meaning» (Aquinas 1269).

It is obvious that Aquinas addition regarding feeling and intellectual activities within the material world is valid only for animals and human beings. In both definitions an important assumption is made: If living objects perform their own activities, like self-nutrition or growths, it means that they are subjects clearly distinguished from the environment. And as subjects they possess a high level of unity. The aspect of unity is important and may be illustrated by the following, perhaps somewhat crude example. If a stone is divided, one obtains two stones with a reduced size. If a dog is divided one gets two pieces of a cadaver or in the best case a stunted dog and some body parts.

In order to find a sufficient reason for the observed unity in living -and also lifeless objects- Aristotle and his followers developed the idea of matter and form, or with Greek terms the hylomorphism. Life is supported by a principle of life, the form (Greek *morphe*, Latin *forma*) which informs a certain matter (Greek *hyle*, Latin *materia*) such that it is a living organism. In the above given comparison between a dog and a stone one could say that information makes the difference, as the form is responsible for something being what it is. The form of living objects Aristotle coined psyche (soul). It is important to note that the form is a metaphysical principle of life, which means in general not an object of reality existing on its own in- or outside the organism. With the hylomorphism Aristotle was able to make a philosophical analysis of change in terms of a remaining principle, the matter and the exchanged one, the form.

It can happen that old concepts with strange names like hylomorphism seem to be outdated. Translated in modern language, however, one finds an unexpected actuality. The form, the principle of life, can be considered as information implemented in a suitable matter, or in computer terms as software implemented in a certain hardware. There is no standalone software. It has always to be implemented in a piece of hardware. For example, a document has to be on a CD, DVD, memory stick or any other hardware medium. On the other hand, hardware always has a certain form, a random bit pattern or only 0's or only 1's. Only combinations of hard- and software occur in the real world, and similarly real objects in our visible world consist of informed matter and never just of standalone matter or standalone forms.

Coming back to living organism one find in the definition above a second essential aspect: living matter exhibits an activity on its own, unexpected when observing an environment that remains at rest. The dog, for example, that is jumping towards the children needs no moving parts in the garden where he was watching. In order to enable this activity there is something new in living matter. It appears as an organism, i.e. a structure containing parts with different functions, 'organs'. To make this possible one needs in the first place a new level of information. The form or information is not only the reason that the living being is as it is, but is also responsible for its characteristic functional behavior and development in time. The latter includes among others growth and restorations after injuries as well as reproduction. In the philosophical analysis based on the A-T approach with only pre-scientific experience one obtains the following fundamental results: There are three levels of life: plants, animals and human beings all with certain characteristics. In the higher levels the lower levels are present. When considering human beings, one finds processes (movements in philosophical terms), which are completely vegetative, for example division of cells during growth. That does not mean that human beings have a set of several forms: vegetative, animal and human. Instead, they have only one single form that includes all levels including the inorganic aspects.

One should also recall that being and form are concepts with a relative, not univocal meaning, they are analogous, transcendental terms, as they can be applied at several levels (Blanchette 2005). Considering a dog, for example, we see a single well-defined being. One can consider also a single hair of this dog and one will find no difference between the hair still rooted in the skin or one just fallen on the ground. Even when the dog is dead the hairs will continue to grow during the first hours after its death. That means that a dog should be considered as a unity but when focusing on specific aspects one seems to deal only with an agglomeration of parts. Applying the paradigm software and hardware for the traditional terms form and matter, one could say that the software, i.e. the information implemented in that dog, is built up of hierarchical building blocks. These building blocks are already used in the lower levels, like the information needed for cell division, and can be studied separately by scientific means, like bio-chemistry, gen-technology etc. It is obvious that these particular sciences alone will never give a total view of a dog due to their willingly restricted approach.

Before Aristotle another philosophical system tried to make change intelligible. The Atomists (Leucippus and Democritus) reduced the variety of beings and their change to the geometrical arrangement of identical parts, the atoms. That view can explain the fact that in all things one finds the same building blocks, but fails completely to provide intelligibility to the strong unity in living beings. One could say that the Atomists propose a still rudimentary version of hylomorphism. What later with Aristotle would be metaphysical principles, matter and form, are still considered as beings on its own and geometry. The atoms, the material in A-T terms, are already complete beings and the form is reduced to continuously varying geometrical arrangements. The ideas of the atomists were taken up in the 17$^{th}$-19$^{th}$ centuries and are often implicitly assumed by reductionists. In their most radical view all beings, including animals and man, are nothing but agglomerations of atoms with a certain arrangement.

In the seventeenth century Descartes considered living beings as complex machines. Regarding man he writes that the bodily movement is not more or less than the movements of a watch or other automaton (Descartes 1632). This view, of course, first appears quite shocking, but expresses an important observation. If one analyses the human body, one will eventually find nothing else than what can be found in the inorganic world. The laws of physics are valid in engineering as well as in biological systems. This observation, of course, is not able to explain the full spectrum of biological phenomena but excludes certain kinds of vitalism (Bechtel and Richardson 1998) that claim the existence of a non-localized substance responsible for biological functions.

Currently the main stream of biologists assumes, at least implicitly, a light or strong version of reductionism. Plants, animals and even human beings are nothing but an arrangement of atoms or molecules in a certain order. When explaining the occurrence of order they follow Darwin's line of reasoning regarding evolution. They state that the changes and the occurrence of order are exclusively based on randomness or chance and subsequent natural selection. Here a comment could be given on this statement, as it belongs to the fundamentals of physics and, more precisely, QM that the occurrence of chance in a particular event cannot be detected by any scientific experiment (Driessen 2010). The conclusion, therefore, about randomness or alternatively any indication of willful design cannot be drawn by a pure scientific analysis, but only by reflection on scientific results within the realm of philosophy and metaphysics.

In summary one can say that most of the views presented emphasize important aspects of life. While doing so, however, other aspects become underexposed or sometimes even completely unintelligible. The hylomorphism of Aristotle, as an intermediate approach between vitalism and strict materialistic reductionism appears as a good starting point for further analysis of scientific experience.

## 3. Need of a system approach in science

In the previous section the high degree of unity in living objects in contrast to inorganic ones has been stressed. Unity means that one should first consider the whole or system and only thereafter the parts or components. The whole is more than just the sum of the parts, a view shared largely in modern physics (Healey 2009). The question arises about the nature of the whole. Are the fundamental laws governing the behavior of the parts still valid when the parts are integrated in a system? It seems proven experimentally that atoms or molecules from, let's say a dog, react in exactly the same way as can be expected from standard chemistry. In that sense the reductionist approach is correct that living matter is governed by the same fundamental laws as lifeless matter. But how to take into account unity and order or, in philosophical terms, the form, i.e. the information implemented in living object? In his landmark article, P.W. Anderson (1972), distinguishes between the road from the complex system to the parts (reductionism) and the other road from parts to the whole (constructionism). He writes:

«The ability to reduce everything to simple fundamental laws does not imply the ability to start from those laws and reconstruct the universe».

He accepts the thesis of the reductionist, but without necessarily excluding that the whole is subjected to natural laws that appear only at the higher level. In a few examples taken from his specialism, namely many body physics, he demonstrates that when increasing scale and complexity a shift from quantitative to qualitative differentiation may occur. That means that the behavior of the system cannot be understood from an extrapolation of the parts.

It is remarkable that most of the examples of Anderson involve QM. The question arises why this theory that started about a century ago, is so important for our subject. In popular discussions on QM the statistical character of the predictions is always mentioned and, relatedly, the Heisenberg uncertainty relation. Also the particle-wave dualism with the resulting complementary descriptions is given special attention. Another equally important aspect, is less emphasized, namely that in QM there are changes that go beyond the geometrical rearrangement of particles. In philosophy one speaks of qualitative changes, if not only quantitative aspects like size and geometrical arrangement are involved but also quality, i.e. a transformation to a different kind of thing (for a philosophical discussion of qualitative properties in nature see Selvaggi (1996)). A qualitative change means that the system or the new thing is something completely different from the starting constituent parts. The occurrence of qualitative changes in QM is explained by Herwig Schopper, the former director of CERN, in a lecture to a general audience. He describes the LEP experiment where electrons and positrons

(a positron is the antimatter of an electron) are accelerated before they collide at high energies:

«Let me repeat with a clear-cut example what is going to happen: We fire two strawberries at each other, in this case a strawberry and an anti-strawberry. Everybody expects that one gets strawberry mush. But this does not happen. In the collision a high concentration of energy is formed and what originates in the next moment is something completely new, namely again fruits. These are completely new fruits, larger and heavier than the original strawberries». (Schopper 1991).

In the following some basic laws of QM are examined that reveal quantum weirdness but simultaneously allow an improved elaboration of the related philosophical concepts. For our argumentation It would be interesting to have the possibility to perform an experiment where the outcome is related to the whole and not to the constituent parts. Let's start with considering particles with a certain mass $m$. According to de Broglie, a particle can be considered as a wave and vice-versa. The so-called de Broglie wavelength $\lambda$ of a particle travelling at speed $v$ is:

$$\lambda = \frac{h}{mv} \qquad (1)$$

where $h$ is the Planck constant. Consider now that the particle is composed of two identical parts, for example a $H_2$ molecule that is composed of two H atoms. Theory tells us that with identical $v$ one gets for the de Broglie wavelength: $\lambda_H = 2\lambda_{H2}$. If the molecule would be the same as just two H atoms, the wavelength would be $\lambda_H$, only the intensity of the wave would be different. That means that in theory a hydrogen molecule is treated as a system, not as a compound of two parts.

What about the experiment? It is known that wave properties, like the wavelength, can experimentally relatively easily be determined by an interference set-up. The Young double-slit arrangement is the most simple arrangement. Particles are emitted by a source at equal speed in front of the two slits and are detected at the screen behind the slits with a movable particle detector or a detector array. In a classical approach with a particle source one would expect after some time two spots at the screen: a blurred image of the two slits. QM reveals the wave properties by displaying a fringe pattern on the screen. In the above given example with hydrogen atoms or hydrogen molecules one gets a periodic local maximum for the molecules exactly at two sets of positions, one where the atom pattern has a maximum and, quite surprisingly, another where it has a minimum. That means that half of the $H_2$ molecules are detected at positions where the count-rate for H atoms would be zero. One therefore may conclude that in the $H_2$ experiment no hydrogen atoms are involved. The double slit experiment distinguishes between the system ($H_2$) and the constituent parts (H). Any purely geometrical arrangement of two H atoms would lead to the same pattern as the single atom.

Meanwhile interference experiments have been done with Helium (atomic mass unit a.m.u. 4), buckyballs (a.m.u. 720) (Nairz et al. 2003) and large organic molecules (a.m.u. up to 6910) (Gerlich et al. 2011). All experiments show the expected fringes for particles with the given mass. This demonstrates that at least in these type of experiments the relatively large particles are treated by nature as systems. Coming back to Anderson who assumes that the reductionist road is always open, one expects that the constituent parts can be made visible by other types of experiment. Only the energy scale should then be extended to compensate in our example for the binding energy of the hydrogen molecule.

Looking only for the mass of a system is a convenient but nevertheless quite superficial way of analysis. More interesting is to study the functional or system behavior that leads to more complex interaction with the environment. In this respect the experimental study of light-harvesting of marine algae is a real break-through. Collini et al. (2010) provide evidence that a complex biological protein exhibit quantum-coherent sharing of electronic excitation across its width of 5 nm. This suggests that distant molecules within the protein are 'wired' together by quantum coherence for more efficient light-harvesting. It proves that a compound with an estimated a.m.u. exceeding $10^6$ is acting coherently. Two other examples are reported in the literature where there is evidence for QM enhancing the functional behavior of biological systems. One is the unexpected ability of some birds, among others the robin, to orient themselves by detecting the gradient of the magnetic field (Rodgers and Hore, 2009). The other is the capacity of fruit flies (Drosophila) to smell by detecting the vibrational spectrum of odorant molecules (Franco et al. 2011). The functional behavior in all these three cases provides evidence that QM is involved, for a detailed discussion see Lambert et al. (2013).

It is appropriate to cite once again Anderson:

«The constructionist hypothesis breaks down when confronted with the twin difficulties of scale and complexity. The behavior of large and complex aggregates of elementary particles, as it turns out, is not to be understood in terms of a simple extrapolation of the properties of a few particles. Instead, at each level of complexity entirely new properties appear. […] At each stage entirely new laws, concepts and generalizations are necessary». (Anderson 1972).

What is the new what one can find in the higher level system? Anderson mentions new laws and concepts; the philosopher Herman Dooyeweerd (1936) speaks of a structure of modal aspects of reality with new law-spheres. In the A-T tradition one would say that at the higher level a new formal cause is acting. And the formal cause is related to information. One could say that the new laws at a higher level, mentioned by Anderson, bring additional information to the system. And this is exactly what is needed to enable the enhanced functional complexity.

A-T philosophy makes use of an important subtlety when dealing with the reality of objects. Besides being real or not real, they can also potentially or virtually be real. To give an example, the beautiful designed apartment in the imagination of an architect is not real, but can perhaps already be seen on the computer screen in virtual reality. Only after many hours of work of skilled craftsmen it becomes a reality. Being real or being an object of reality is a serious classification, as it means among others that it is by definition a being without any internal contradiction. That is why experiments have always the last word when comparing with theories, models and computer simulations. The opposite way is not necessarily true, being without internal contradiction does not imply that the object in question does really exist.

With this introduction one can now examine the properties of the constituent parts in a whole. Let's start with the Heisenberg uncertainty principle and apply it to the hydrogen atom, consisting of a proton and an electron. It states that the product of the uncertainty in position $\Delta x$ times the uncertainty in momentum $\Delta p$ has a minimum value

$$\Delta x \, \Delta p \geq h/(4\pi) \qquad (2)$$

where h is the Planck constant. It appears that the radial position of the electron in the ground state has an uncertainty of about half the radius of the electron orbit in the Bohr model -for details see standard textbooks on physics like Cutnell and Johnson (2007). We end up with the surprising result that the electron is not localized in a fixed radial position, as assumed in the Bohr model. The laws of QM provide only the probability to find the electron at a certain distance of the proton.

The situation is even more complex. In the Bohr picture the electron should circle around the proton with a resulting magnetic moment. This indeed is experimentally confirmed for the excited state. But a circulating charge should emit radiation that would slow down immediately the speed of the electron. But in contradiction to Maxwell's equations this has never been observed (Bohr was aware of this inconsistency). Obviously the electron inside the hydrogen atom does not behave as a common electron. We therefore have to conclude that it is not possible to observe theoretically or experimentally the detailed properties of the electron inside the hydrogen atom. The philosopher could say it is only virtually or potentially present; in that case the contradictory properties (indeed a rotating charge plus a magnetic moment, but no radiation damping) are no problem. What is real is the whole or the system, the hydrogen atom with a certain mass, charge distribution, magnetic moment, etc. With sufficiently strong causes it is possible to obtain a real electron from a hydrogen atom. But doing so, the atom will be destroyed, the system has been reduced to its parts.

Heisenberg explicitly connects the probabilistic character of the electronic state with Aristotelian philosophy. He writes:

«One might perhaps call it an objective tendency or possibility, a "potentia" in the sense of Aristotelian philosophy. In fact, I believe that the language actually used by physicists when they speak about atomic events produces in their minds similar notions as the concept "potentia." So the physicists have gradually become accustomed to considering the electronic orbits, etc., not as reality but rather as a kind of "potentia."» (Heisenberg 2007, pp. 154-155).

More generally one can state that once the compound, the whole or the system is established, the constituent parts are only virtually present. A comparison with mechanics could illustrate this. Consider the parts of a classical clock. It would be astonishing that a gear, once assembled, would become a diffuse object, a kind of sponge, and appears as soft as butter. After disassembling, however, it would be again hardened steel in its original shape. QM allows such a weird behavior, i.e. the system may have properties which could not been expected when considering the constituent parts on its own.

A system intensively studied after the appearance of Anderson's Science paper are entangled photons. These are generated, for example, by down conversion in a nonlinear optical crystal. In Bell-type experiments (Bell 1997) the polarization states of the individual photons are measured and compared with theoretical predictions. Any theory assuming that the polarization states of the individual photons before measurements are fixed but nevertheless unknown, is in contradiction with experiments. With other words, considering entangled photons as just a pair of photons with properties single photons normally exhibit, will not work. It means that physical reality cannot be described mathematically by a product of wave functions each related to one of the parts. Instead certain combinations of wave functions, each one excluding the other, have to be taken into account to allow an adequate description of the system, for details see (Healey 2009).

Using A-T terminology for the case of entanglement one could say that two entangled photons are a system, where the parts – the individual photons – exist only virtually, but not actually. Only the measuring process, or collapse of the wave function, converts the whole in two actually existing parts, the two photons. In that case the superposition state is changed to the product of the wave functions of the individual parts. Heisenberg states in a similar context:

«The probability wave of Bohr, Kramers, Slater […] was a quantitative version of the old concept of "potentia" in Aristotelian philosophy. It introduced something standing in the middle between the idea of an event and the actual event, a strange kind of physical reality just in the middle between possibility and reality» (Heisenberg 2007).

There are more examples where specific QM states indicate a system that cannot be explained alone by the properties of its parts. In superconductivity two electrons (fermions) are coupled to form a Cooper pair, which as bosons obey a completely different quantum statistics in comparison with the of the individual electrons. These bosons at sufficient low temperature may undergo eventually a Bose-Einstein condensation leading to a macroscopic quantum state, the superconducting state. Superfluidity of helium is another example of Bose-Einstein condensation already mentioned in Anderson's article, for more details see (Amico et al. 2008).

It is not the place to discuss in detail these examples. But one point should be made: Even in relatively simple systems in lifeless matter, e.g. two electrons that form a pair, one observes unexpected QM behavior. For the pairs (and not the individual electrons) are responsible for a completely new phenomenon, superconductivity, that could not be predicted from the individual parts (the electrons). On the other hand, any analysis of the system of two electrons in terms of the parts (the individual electrons) reveals only the expected behavior as predicted by reductionism, i.e. the normal electrical conductivity.

## 4. Discussion and concluding remarks

In the preceding section it has been tried to show that QM allows a view on reality different of the 19$^{th}$ century view of classical physics. The most remarkable is, in my opinion, the evidence that there are qualitative changes in nature. These changes occur where a new form or in other words, new information is implemented in the system. That means that the systems in nature cannot be reduced to a more or less complex arrangement of the constituent parts. In A-T terms of the four aspects of causality one could say that the formal aspects have to be considered. Focusing on the materials aspects, in this case the properties of the constitutive parts, is not sufficient. The main reason for this is that the constituent parts are only virtually present in the whole, as discussed above in the case of the hydrogen atom. Any notion of a localized electron in the atom leads to a contradiction in theory as well as in experiment. The same happens in the double slit experiment with light, atoms or molecules. If one tries to identify experimentally or theoretically through which of the two slits the object has passed, the interference pattern disappears. That means that the system of double slits would have been substantially changed.

Is this such a weird situation? For the Greek philosophy of Aristotle and most other people it has been and is also now evident that there are changes beyond the pure rearrangement of parts. An example taken of the field of art may illustrate this more clearly. Contemplating the masterpieces of painters and sculptors of all times one recognizes in the faces of portrayed persons not only details of form, colors, eyes, mouth, skin and hair but also beauty, passion and a kind of summary of the personal history of that person. In the originals, the faces of living persons, an enormous information is stored far beyond that could be expected from geometrical arrangements of biological tissues.

The arguments up to now are valid for both, living and lifeless matter. What can be said about the field of biology, the science that deals with life in all its material appearances? In the introduction and first section we spoke about the mystery of life and the clear distinction that both classical philosophy and common people make between living and non-living objects. It would be worthwhile to assume that there is an ontological basis for this distinction. The distinction is not in the parts one obtains after disassembling the whole. These are in common with those obtained from disassembling lifeless matter. Instead one should look for a higher-level organization made possible by a qualitative change connected with increased information. The above mentioned experimental results on enhanced light-harvesting, improved aviation and smelling provide evidence for the correctness of the system approach. These are specific subsystems studied in biology. If one now consider living organism from virus to human beings, could it be that what we call life is a system organized in a kind of macroscopic super-entanglement? Vedral (2011) arrives at a similar assumption. He states:

«Other experiments scale up this basic idea, so that huge numbers of atoms become entangled and enter states that classical physics would deem impossible. And if solids can be entangled even when they are large and warm, it takes only a small leap of imagination to ask whether the same might be true of a very special kind of large, warm system: life».

In the light of the preceding a few suggestions can be made for biologists and philosophers. In classical philosophy there is a well-known adage, *agere sequitur esse*, which can be considered as a variant of the principle of causality. Translated literally as *action follows being* it expresses the idea that if something happens, then there should be an ontological basis, with other words, effects arise only from existing causes. This principle lies at the basis of any scientific work. Bearing this in mind biologists should dare to use phantasy and trust more in the creativity of nature. With the words of Seth Lloyd one may state: *biology has a knack for using what works* (quoted in Ball, 2011). If one observes enhanced efficiency in light harvesting of certain biological nanostructures that goes beyond the capabilities of the constituent parts, then there should be an ontological basis, i.e. a system responsible for this. The same can be said for other scientific results. If birds can detect very small magnetic signals in order to orient themselves during flight or when fruit flies differentiate between molecules with different vibration levels but otherwise identical chemical properties then there should be a complex system responsible for this. In all three cases QM seems to be the only explanation as discussed above. The classical ball and spring models for molecules cannot account for the extraordinary performance of the system. Instead one has to use a system approach where the whole is qualitatively different from the aggregation of the constituent parts.

Immediately another question arises about an important point of discussion in evolution. If one encounters evidence for evolutionary paths with essentially zero probability (Meyer 2009), (Lloyd 2002) within the probabilistic resources of the universe, should one then not look for systems which allow a much more efficient route to complexity than a blind trial and error approach? In the light harvesting experiments one encounters enhanced efficiency of transport of excitons towards the absorption centers due to quantum interference. Ball explains it as follows:

«In fact, they (the coherent quantum waves) could simultaneously explore a multitude of possible options, and automatically select the most efficient path to the reaction centre» (Ball, 2011).

Could it not be that the additional information available in higher level biological systems allows also a much more efficient route to mutations with improved adaptations to the environment? In both cases QM could be the means how nature provides additional information to the system so that steps to a profitable situation of the system are less random or blind than one would expect from the statistical properties of the parts. The origin of additional information would be due to the natural laws governing the higher level system. In a certain sense saying that at a certain step qualitative changes occur is the same as saying that nature is providing new information. In this way a hierarchy of complexity can be obtained in accordance with experience, see, e.g. (Mitchell, 2009), and the A-T view on increasingly higher level of formal causality.

A recommendation could also be made to philosophers. In classical philosophy the available experience at a given culture was used to establish the philosophical system. Would it not be worthwhile to use the immense increase in knowledge to work on methods for performing a consistency check on the different philosophical systems? For example, in a philosophical framework that ignores qualitative changes QM would be not intelligible. Also Kant idealism has great problems, as is pointed out in a study on relativity and QM by Mittelstaedt (1975). But in order to do this, an open dialogue between scientists and philosophers is needed as the scientific evidence is often written in a formal language incomprehensible to outsiders. On the other side one finds among scientists often only elementary knowledge of philosophy and a certain unfamiliarity with abstract philosophical reasoning.

There is an additional point to be made. Philosophy has the ambition to reach truth and attain wisdom that means to provide answers to the deepest questions humans can ask. What mostly can be seen instead is a very high level of erudition together with a certain reluctance to make choices and acknowledge truth. In science a theory is abandoned when it is falsified by new insights. It would be a matter of wisdom to put less confidence on philosophical systems that do not surpass a consistency check not only with common sense knowledge but also with sophisticated scientific results obtained in science or mathematics (Driessen, 2005). Hawking makes a challenging statement about philosophy in his book *The Grand Design* (Hawking and Mlodinow, 2010):

«Traditionally these are questions for philosophy, but philosophy is dead. Philosophy has not kept up with modern developments in science, particularly physics».

The reproach addressed to the philosophers is perhaps exaggerated but lack of life, i.e. disintegration is surely observable. It is modesty or perhaps a pessimistic view on the capacity of the human ratio that leads to an auto-limitation to particular aspects of knowledge instead of offering a coherent philosophical reflection on reality. In order to include the results of modern science in the philosophical reflection, more interaction between scientist and philosophers is clearly needed. In that interdisciplinary effort, both philosophers and scientists could profit enormously ending up eventually with an satisfying answer to the question, "What is life?".

## Literature

- Abbot, D., Davies, Paul C.W. and Pati A.K. (Eds.), *Quantum Aspects of Life,* Imperial College Press, 2008.

Amico, L., Fazio, R., Osterloh, A. and Vedral, V., *Entanglement in Many Body Systems*, arXiv:quant-ph/0703044v3, 9 May 2008.
Anderson, P.W., *More is different*, «Science», New Series, Vol. 177, No. 4047, 1972, pp. 393-396.
Aquinas, Thomas, [1269], *In II De anima*, lectio 1, n. 219.
Arecchi, T.F., *A critical Approach to Complexity and Self Organization*, «Studies in Science & Theology», 1 (1993), Labor et fides, pp. 5-29.
Aristotle, *De anima*, book II, The Internet Classic Archive, http://classics.mit.edu/Aristotle/soul.2.ii.html.
Ball, P., *The dawn of quantum biology*, «Nature», 474, 2011, pp. 272-274.
Bechtel, W. and Richardson, R.C., *Vitalism*, in E. Craig (Ed.), *Routledge Encyclopedia of Philosophy*, Routledge, London 1998.
Bell, J.S., *Indeterminism and nonlocality*, in A. Driessen and A. Suarez (eds.), *Mathematical Undecidability, Quantum Nonlocality and the Question of the Existence of God*, Springer, 1997, pp. 83-100.
Blanchette, O., *Analogy and the Transcendental Properties of Being as the Key to Metaphysical Science*, «The Saint Anselm Journal», 2/2 (Spring 2005).
Brigandt, I. and Love, A.C., *Reductionism in Biology*, «The Stanford Encyclopedia of Philosophy», Summer 2012 Edition, E.N. Zalta (ed.), http://plato.stanford.edu/archives/sum2012/entries/reduction-biology.
Collini, E., Wong, C.Y., Wilk, K.E., Curmi, P.M.G, Brumer, P. and Scholes, G., *Coherently wired light-harvesting in photosynthetic marine algae at ambient temperature*, «Nature», 463, 2010, pp. 644-647.
Cutnell, J.D. and Johnson, K.W., *Physics*, John Wiley & Sons, 2007.
Davies, Paul C.W, 2004: *Does quantum mechanics play a non-trivial role in Life?*, «BioSystems», 78, 2004, pp. 69-79. doi: 10.1016/j.biosystems.2004.07.001.
Del Re, G., *Complexity, Organization, Information*, «Studies in Science & Theology», 1 (1993), Labor et fides, pp. 83-92.
Descartes, R., *Traité de l'homme*, [1632]. *Oeuvres* VI (Adam et Tannery), pp. 58-59.
Dooyeweerd, H., *De wijsbegeerte der wetsidee*, H.J. Paris, Amsterdam 1936.
Driessen, A., *The question of the existence of God in the book of Stephen Hawking "A Brief History of Time"*, «Acta Philosophica», 4, 1995, pp. 83-93.
Idem, *Philosophical consequences of the Gödel Theorem*, in E. Martikainen (ed.), *Human approaches to the universe*, Luther-Agricola-Society, 2005, pp. 66-74.
Idem, *Evolutie bekeken met de bril van een natuurkundige*, in A. Driessen and G. Nienhuis (eds.), *Evolutie: wetenschappelijk model of seculier geloof*, Kampen, Kok 2010, pp. 33-45.
Franco, M.I., Turin, L, Mershin, A. and Skoulakis, E.M.C., 2011: *Molecular vibration-sensing component in Drosophila melanogaster olfaction*, PNAS, 108, 2011, pp. 3797-3802, www.pnas.org/cgi/doi/10.1073/pnas.1012293108.
Eigen, M., *Wie entsteht Information? Prinzipien der Selbstorganisation in der Biologie*, «Ber. Bunsengesellschaft f. phys. Chem.», 80 (1976), pp. 1059-1081.
Gerlich, S., Eibenberger S., Tomandl, M., Nimmrichter, S., Hornberger, K., Fagan, P.J., Tüxen, J., Mayor, M. and Arndt, M., *Quantum interference of large organic molecules*, «Nature Commun.», 2 (2011), 263 doi: 10.1038/ncomms1263.

Hawking, S., *A Brief History of Time, from the Big Bang to Black Holes*, Bantam Books, New York 1988.
Hawking, S. and Mlodinow, L., *The grand design,* Bantam books, New York 2010.
Healey, R., *Holism and Nonseparability in Physics*, in E.N. Zalta (ed.), *The Stanford Encyclopedia of Philosophy* (Spring 2009 Edition), http://plato.stanford.edu/archives/spr2009/entries/physics-holism.
Heisenberg, W., *Physics and Philosophy*, Harper Perennial Modern Classics edition, 2007, cited by E. Frazer on his blog: *Heisenberg on act and potency*.
Hoenen, P., *Philosophie der Anorganische Natuur*, Standaard Boekhandel, Antwerpen 1947.
Lambert, N., Chen, Y.N., Cheng, Y.C., Li, C.H., Chen, G.Y. and Nori, F., *Quantum biology*, «Nature Physics», 9 (2013), pp. 10-18.
Lloyd, S., *Computational capacity of the universe*, «Phys. Rev. Letters», 88, 2002, pp. 7901-7904.
Idem, *Quantum coherence in biological systems*, J. Phys., Conf. Ser. 302, 012037, 2011.
Meyer, S.C., *Signature in the cell*, HarperOne, New York 2009, chapter 10.
Mitchell, M., *Complexity, a guided tour*, Oxford University Press, Oxford 2009.
Mittelstaedt, P., *Philosophical problems of modern physics,* Springer Netherlands 1975.
Nairz, O., Arndt, A. and Zeilinger, A., *Quantum interference experiments with large molecules,* «American Journal of Physics.», 71/4 (2003), pp. 319-325.
Panichayangangkoon, G., Hayes, D., Fransted, K.A., Caram, J.R., Harel, E., Wen, J., Blankenship, R.E., Engel, G.S., *Long-lived quantum coherence in photosynthetic complexes at physiological temperature*, «PNAS - Proceedings of the National Academy of Science», 107, 2010, pp. 12766-12770.
Rodgers, C.T. and Hore, P.J., *Chemical magnetoreception in birds: The radical pair mechanism*, «PNAS - Proceedings of the National Academy of Science», 106, 2009, pp. 353-360; see www.pnas.org_cgi_doi_10.1073_pnas.0711968106.
Romero-Isart, O., Juan, M.L., Quidant, R. and Cirac, J.I., *Toward quantum superposition of living organisms*, «New Journal of Physics.» 12, 2010, 0333015.
Schopper, H., *Was heißt Materie? Beiträge der Elementarteilchenphysik zum Weltverständnis*, in H. Thomas (Ed), *Naturherrschaft*, Busse Seewald, Herford 1991, pp. 11-34.
Selvaggi, F., *Filosofia del mondo*, 2nd edition, Editrice Pontificia Università Gregoriana, Roma 1996, Chapter XV, pp. 359-383.
Soler Gil, F.J., *Aristoteles in der Quantenwelt*, Peter Lang, Frankfurt a.M. 2003.
Vedral, V., *Living in an Quantum World*, «Scientific American», June 2011, pp. 38-43.
Weber, B., Life, *The Stanford Encyclopedia of Philosophy*, (Winter 2011 Edition), Edward N. Zalta (ed.), 2011, http://plato.stanford.edu/archives/win2011/entries/life.

Abstract: *The rapidly increasing interest in the quantum properties of living matter stimulates a discussion of the fundamental properties of life as well as quantum mechanics. In this discussion often concepts are used that originate in philosophy and ask for a philosophical analysis. In the present work the classic philosophical tradition based on Aristotle and Aquinas is employed which surprisingly is able to shed light on important aspects. Especially* *one could mention the high degree of unity in living objects and the occurrence of thorough qualitative changes. The latter are outside the scope of classical physics where changes are restricted to geometrical rearrangement of microscopic particles. A challenging approach is used in the philosophical analysis as the empirical evidence is not taken from everyday life but from 20th century science (quantum mechanics) and recent results in the field of quantum biology. In the discussion it is argued that quantum entanglement is possibly related to the occurrence of life. Finally it is recommended that scientists and philosophers should be open for dialogue that could enrich both. Scientists could redirect their investigation, as paradigm shifts like the one originating from philosophical evaluation of quantum mechanics give new insight about the relation between the whole en the parts. Whereas philosophers could use scientific results as a consistency check for their philosophical framework for understanding reality.*

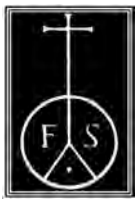